\documentstyle[12pt,aasms4]{article}
\begin{document}
\title{THE NEAR-INFRARED SPECTRUM OF THE BROWN DWARF GLIESE 229B}

\author{T.R. Geballe}
\affil{Joint Astronomy Centre, 660 N. A'ohoku Pl., Hilo, HI  96720}
\authoremail{tom@jach.hawaii.edu}

\author{S.R. Kulkarni}
\affil{Palomar Observatory 105-24, California Institute of Technology, 
Pasadena, CA 91125}
\authoremail{srk@astro.caltech.edu}

\author{Charles E. Woodward$^{1,2}$}
\affil{University of Wyoming, Department of Physics \& Astronomy,
 Laramie, WY 82071-3905}
\authoremail{chelsea@wapiti.uwyo.edu}

\author{G.C. Sloan$^{1}$}
\affil{NASA Ames Research Center, MS 245-6, Moffett Field, CA
94035-1000}
\authoremail{sloan@ssa1.arc.nasa.gov}

\vskip 4.0 truecm

\noindent $^{1}$ Visiting Astronomer, United Kingdom Infrared Telescope.

\noindent $^{2}$ NSF Presidential Faculty Fellow.

\begin{abstract}
A medium resolution $1.0-2.5~\mu$m spectrum of the brown dwarf, Gliese~229B
has been obtained using CGS4 on UKIRT. In addition to the broad spectral
structure seen in earlier low resolution observations, the new spectrum
reveals a large number of absorption lines, many of which can be identified
with water vapor.  Water and methane are both shown to be strong absorbers in
the near-infrared spectrum of the object. Several spectral features in Gl~229B
that are attributable to methane match ones seen in reflection in the giant
outer planets and, in particular, Titan. 

\end{abstract}

\keywords{infrared: general --- infrared: stars --- infrared: lines and bands
--- stars: individual (Gl~229B) --- stars: low mass, brown dwarfs} 

\section{INTRODUCTION}

Recently, Nakajima et al. (1995) reported the first detection of a cool brown 
dwarf, Gliese~229B (hereafter Gl~229B), a proper motion companion of 
Gliese~229A. Oppenheimer et al. (1995) obtained a low resolution ($\lambda 
/ \Delta\lambda \approx 150$) near-infrared ($1-2.5~\mu$m) spectrum
of Gl~229B and found a number of strong absorption bands.  Similar absorption
bands are seen in the spectrum of Jupiter and are attributed to methane. 
Methane is not seen in stars.  Tsuji, Ohnaka \& Aoki (1995) conclude that
methane will be seen only in objects cooler than about 1800~K, lower than the
effective temperatures of even the least massive stars (Burrows \& Liebert
1993). From the measured broad-band spectrum of Gl~229B and assuming a radius
equal to that of Jupiter, Matthews et al. (1996) infer T$_{\rm eff} = 900$~K
and a bolometric luminosity of $6.4 \times 10^{-6}~L_{\odot}$. The low
luminosity and the presence of methane in the photosphere require that
Gl~229B is a cool brown dwarf. 

Here we present a new $1.0-2.5~\mu$m spectrum of Gl~229B with significantly
higher resolution and signal-to-noise ratio than the original spectrum
presented by Oppenheimer et al. (1995).  The new data provide a considerably
more detailed view of Gl~229B. We compare our spectrum to spectra of the
Jovian planets and Titan which are known to show strong absorption features
due to methane in the near-infrared. 

\section{OBSERVATIONS AND DATA REDUCTION} 

We obtained spectra of Gl~229B in the $0.99-2.52~\mu$m interval at the United
Kingdom 3.8~m Infrared Telescope (UKIRT) on 1995 October 28-29 UT and
December 12-13 UT, using the facility spectrometer CGS4 (Mountain et al. 
1990). The 75 l/mm grating and the 150 mm focal length camera optics in CGS4
imaged a slit $90^{\prime\prime}$ long onto a 256$\times$256 InSb array. 
Each pixel spanned 1$\farcs$23, and the slit was one pixel wide.  Table 1
provides details of the observations. The resolution of CGS4 was approximately
390$\times$$\lambda$($\mu$m) in first order and 780$\times$$\lambda$($\mu$m)
in second order.  To calibrate the spectra, we observed HR 1849 (A0V, V=5.55)
each night just before observing Gl~229B. 

Gl~229B is about $8^{\prime\prime}$  from Gl~229A. In the J, H, and K bands
Gl~229A outshines Gl~229B by approximately 10~magnitudes.  At UKIRT the halo
of diffracted and scattered light from Gl~229A contributed a large amount of 
radiation to the array rows containing the Gl~229B spectrum.  In order to 
minimize the difficulties of removing this background, we observed with the 
slit of CGS4 perpendicular to the line joining the two sources on 
the sky and obtained alternate spectra of Gl~229B by nodding the telescope 10 
rows along the slit. Subtracting these alternate spectral images removed much 
of the halo, allowing Gl~229B to be detected clearly at nearly all 
wavelengths, typically at greater strength than the residual halo from the 
primary.  We extracted the spectrum of Gl~229B using standard infrared 
reduction procedures to correct for curvature of the spectra
and to remove residual background. Spectra of HR 1849 were utilized to
correct for atmospheric and instrumental absorption, after editing them to
remove atomic hydrogen absorption lines (with the exception of
Pa $\alpha$ at 1.875~$\mu$m). 

Figure 1 presents the complete spectrum of Gl~229B from 0.99 to 2.52 $\mu$m (a 
numerical file of this spectrum is available from the authors upon request).  
A more expanded version of the spectrum is provided in Fig. 2. To produce 
the spectrum, we slightly smoothed and rebinned the observed 
spectra in steps of 0.0005 $\mu$m for the second order segments and 
0.0010~$\mu$m for the first order segments.  The smoothing lowers the 
spectral resolving power by $\approx 10$\% from the actual instrumental 
resolution cited above. In the region of overlap between the two orders, we 
chose to use the first order spectrum beyond 1.585~$\mu$m.  We scaled the 
individual segments (by factors ranging from 0.7 to 1.2) to match the spectra 
in overlapping sections and at 1.585~$\mu$m.  The average scaling factor was 
unity, but we scaled the final spectrum by a factor of 0.76 to more 
accurately match (to within about 0.1~mag) the recent Palomar photometry of 
Gl~229B (J=14.2, H=14.3, Ks=14.3, K=14.4) as reported by Matthews et al. 
(1996a). 

The noise in the CGS4 spectrum varies considerably with wavelength, as 
illustrated in Fig. 2, due to large variations in atmospheric transmission
and in the background (caused by both thermal and OH emissions). In
particular, in the spectral regions dominated by strong telluric absorption
bands ($1.12-1.14~\mu$m, $1.36-1.42~\mu$m, $1.81-1.93~\mu$m, and
$2.48-2.52~\mu$m), apparent spectral features may not be real. In these
spectral regions the data probably are only useful for estimating the average
flux level. 

\begin{deluxetable}{lllll}
\tablewidth{0pt}
\tablenum{1}
\tablecaption{Observation Log}
\tablehead{
  \colhead{Date (UT)} & \colhead{Grating order} & 
  \colhead{$\lambda$($\mu$m)} & \colhead{Int. Time} & 
  \colhead{Conditions}
}
\startdata
1995 Dec. 12 & second & 0.99-1.33 & 18 m & photometric \nl
1995 Dec. 12 & second & 1.27-1.61 & 24 m & photometric \nl
1995 Oct. 29 & first  & 1.48-2.15 & 24 m & photometric \nl
1995 Oct. 28 & first  & 1.85-2.52 & 30 m & partly cloudy \nl
1995 Dec. 13 & first  & 1.85-2.52 & 36 m & photometric \nl
\enddata
\end{deluxetable}

\section{DESCRIPTION OF THE INFRARED SPECTRUM}

The new spectrum of Gl~229B is consistent with the lower resolution
spectrum of Oppenheimer et al. (1995).  However, the latter spectrum was 
incorrectly normalized, as noted by Matthews et al. (1996a).  Matthews et al. 
(1996b) present a revised spectrum with the proper normalization, with which 
the CGS4 spectrum agrees well.

Two new results are immediately apparent from the new spectrum. First,
emission is seen across the entire IR band from 1.0 to 2.5~$\mu$m.
Specifically, emission is detected through the wavelength intervals of
strong telluric H${_2}$O absorption (see discussion above) and in the
range $2.2-2.5~\mu$m including a notable rise in flux at wavelengths
longward of 2.4~$\mu$m. The data of Oppenheimer et al. (1995) were
restricted to the usual near-infrared windows (Z, J, H, K) and
furthermore were noisy in the range $2.2-2.5~\mu$m.  We emphasize
that, with the possible exception of the wavelength interval
$1.36-1.40~\mu$m, where the detection is marginal, the emission
detected is from Gl~229B and is not contamination by Gl~229A;
crosscuts through the differenced spectral images in these wavelength
intervals show local maxima in the rows of the array corresponding to
the location of Gl~229B. Second, numerous narrow lines are seen (see
Fig.~2). These appear in regions of high emission and on all of the
absorption edges with the exception of the very steep edge at
1.6~$\mu$m. Broader features occur in the $1.6-1.9~\mu$m absorption
trough at about 1.63, 1.67, 1.71, and 1.80~$\mu$m, and at 2.17, and
2.20~$\mu$m. Most of the broad features were seen in the spectrum of
Oppenheimer et al. (1995), albeit at lower resolution.

\section {DISCUSSION}

\subsection{Methane: comparison with Solar System objects}

We have compared the new spectrum of Gl~229B to the solar reflectance 
spectra of three of the giant gaseous planets (Jupiter, Saturn, and Uranus)
and Titan, using the data of Fink \& Larson (1979), and recent CGS4 spectra 
of Titan (Owen, unpublished) and Saturn (Geballe, unpublished).  Methane 
dominates the $1.0-2.5~\mu$m spectra of all of the above outer planets
and Titan, but the degree of domination varies, increasing in the sequence
Jupiter, Saturn, Titan, and Uranus. 

The spectrum of Titan provides the best overall spectral match to that of
Gl~229B and the spectra of the two exhibit several similar details. Most
remarkably, their very sharp cut-offs at 1.61~$\mu$m are essentially
identical (see Fig. 3). The matches with the spectra of the other Solar 
System objects do not appear as precise; for example, the cut-off in Saturn
is shifted by 0.01~$\mu$m to longer wavelength. Between 1.62 and 1.72~$\mu$m 
the wavelengths of some of the absorption maxima in Titan and Gl~229B 
agree.  Both Titan and Gl~229B show absorption bands at $\sim$2.17~$\mu$m and
2.20~$\mu$m (as do Jupiter and Saturn). In Titan each of these features
arises from CH$_{4}$ (Fink \& Larson 1979; C. Griffith, pers. comm.). Thus,
despite the considerably different physical conditions of the CH$_{4}$ in
Gl~229B compared to those in the atmospheres of the giant gaseous planets and
Titan, detailed comparison provides clear confirmation of the strong
influence of methane on the spectrum of Gl~229B.

\subsection {H$_2$O in Gl~229B}

Theoretical model spectra of brown dwarfs with effective temperatures
similar to that of Gl~229B show considerable absorption due to steam
(Marley et al.  1996; Allard et al. 1996). We note, however, that
these model spectra only include low resolution methane opacities
shortward of 1.6~$\mu$m (R.S. Freedman, personal communication). Jones
et al. (1994) have compiled low resolution spectra of the coolest
stars and attribute a broad absorption centered at about 1.4~$\mu$m
and its sharp absorption edge, whose wavelength they report as 1.34
$\mu$m, to H$_2$O. The 1.4~$\mu$m water band also appears strongly in
the above model brown dwarf spectra. Following Jones et al.,
Oppenheimer et al. (1995) identified the absorption edge seen in
Gl~229B at about this wavelength as due to H$_2$O.

In the present spectrum, the absorption edge in Gl~229B is seen to be
centered at about 1.31~$\mu$m. The H$_2$O absorption edge in late M
stars actually occurs at 1.33~$\mu$m (Walker et al. 1996) and is much
steeper than the absorption edge in Gl~229B; indeed it is expected to
strengthen and steepen with decreasing temperature. The methane
absorption edges of Titan and Saturn are centered at 1.30~$\mu$m and
1.31~$\mu$m, respectively.  The work of Jones et al. is restricted to
stars which are cool, but certainly not cool enough to have strong
photospheric methane bands. Thus, confusion between CH$_4$ and H$_2$O does
not arise in the interpretation of stellar spectra. We conclude on the
basis of the low resolution spectra alone that the absorption edge and
trough longward of 1.3~$\mu$m in Gl~229B cannot be attributed
unambiguously to water alone and that methane could contribute
significantly. However, when compared to the spectra of all of the
giant outer planets and Titan, the emission bump at $1.5-1.6~\mu$m is
badly eaten away on its short wavelength side (Fig. 3).  Absorption by
H$_2$O in Gl~229B can explain this systematic difference between
Gl~229B and the giant planets and Titan. The sharpest portion of the
1.3~$\mu$m absorption edge of Gl~229B, at 1.33~$\mu$m, would then be
due to H$_2$O.

Although the low resolution spectra are ambiguous as to the presence
of water vapor, comparison of the new higher resolution spectrum of
Gl~229B with plots of H$_2$O opacity at similar resolution (kindly
supplied by D. Saumon, R.S.  Freedman, and D. Schwenke) demonstrate
its presence clearly. An illustration of this is provided in Fig. 4.
Almost every absorption line in the spectrum of Gl~229B in the
intervals $1.30-1.34~\mu$m, $1.53-1.58~\mu$m, and $1.95-2.07~\mu$m has
a counterpart in the water opacity spectrum. Moreover, there is good
overall correlation between observed and modelled line strengths. The
opacities of the lines in the above wavelength intervals are
relatively low, implying that the depressions in Gl~229B at about
1.4~$\mu$m and 1.9~$\mu$m are indeed due to absorption by H$_2$O,
which has much higher opacity at those wavelengths.

Thus, the new spectrum allows a firm identification of water vapor in the
photosphere of Gl~229B.  It is clear that, as predicted in the models, both 
H$_2$O and CH$_4$ produce strong absorptions in the near-infrared spectrum of 
Gl~229B. However, until the line parameters of methane are better known,
it will be difficult both to disentangle the relative contributions of
methane and water and to search for other absorbers at near-infrared
wavelengths. 

\section{Acknowledgements}

We thank the many individuals responsible for the excellent
construction, maintenance, and operational support of CGS4 and the
staff of UKIRT for its support of these observations. We are grateful
to R.S. Freedman, D. Schwenke, and D. Saumon for providing us with
detailed opacity plots of H$_2$O and CH$_4$ and to T. Owen for
permitting us to publish a portion of the spectrum of Titan. We thank
the referee, F. Allard, for several helpful comments. We also thank
D. Cruikshank, M. Marley, K. Matthews, T. Nakajima, and B. Oppenheimer
for their assistance.  The United Kingdom Infrared Telescope is
operated by the Joint Astronomy Centre on behalf of the U.K. Particle
Physics and Astronomy Research Council.  SRK's research is supported
by Caltech, the US NSF, NASA, and the Packard Foundation. CEW
acknowledges support from the US National Science Foundation.

\vfill\eject

\section{REFERENCES}

\noindent Allard, F., Hauschildt, P.H, Baraffe, I. \& Chabrier, G. 1996,  
ApJ (Letters), in press.

\noindent Burrows, A. \& Liebert J. 1993, Rev. Mod. Phys., 65, 301

\noindent Fink, U., \& Larson, H. P. 1979, ApJ 233, 1021

\noindent Jones, H. R. A., Longmore, A. J., Jamesom, R. F., \& Mountain,
C. M. 1994, MNRAS 267, 413

\noindent Marley, M.S., Saumon, D., Guillot, T., Freedman, R.S., Hubbard, 
W.B., Burrows, A. \& Lunine, J.I. 1996, Science, submitted

\noindent Matthews, K., Nakajima, T., Kulkarni, S., \& Oppenheimer, B. 1996a,
I.A.U. Circ. No. 6280

\noindent Matthews, K., Nakajima, T., Kulkarni, S.R. \& Oppenheimer, B. 1996b,
ApJ, submitted

\noindent Mayor, M., \& Queloz, D. 1995, I.A.U. Circ. No. 6251

\noindent Mountain, C.M., Robertson, D.J., Lee, T.J. \& Wade, R.  1990, in
Instrumentation in Astronomy VII, D.L. Crawford Ed., (Proc. SPIE, 1235), 25

\noindent Nakajima, T., Oppenheimer,. B.R., Kulkarni, S.R., Golimowski, D.A.,
Matthews, K. \& Durrance, S.T. 1995, Nature, 378, 463

\noindent Tsuji, T., Ohnaka, K., Aoki \& W. 1995, in The Bottom of the Main 
Sequence - and Beyond, C.G. Tinney, Ed. (Springer, Berlin), p. 45

\noindent Walker, H.J., Tsikoudi, V., Clayton, C.A., Geballe, T.R., Wooden, 
D.H. \& Butner, H.M. 1996, A\&A, submitted

\section{FIGURE CAPTIONS}

\figcaption[]{Combined 1.0 -- 2.5~\micron \ spectrum of Gl~229B. The intensity
scale is logarithmic; negative values near 1.4~$\mu$m were removed prior to
plotting} 

\figcaption[]{Spectral segments for Gl~229B, covering $0.99-2.52~\mu$m.
Representative noise levels are shown as vertical lines along the bottom of 
each panel.  a. $0.99-1.41~\mu$m.  b. $1.39-1.91~\mu$m.  c. $1.89-2.52~\mu$m.}

\figcaption[]{Spectra of Gl~229B (darker line) and Titan (scaled by a factor
of $\sim$0.003) near the 1.6~$\mu$m methane absorption edge.} 

\figcaption[]{Spectrum of Gl~229B and the opacity of water vapor 
in the $1.95-2.10~\mu$m region.  The opacity is for P~=~1~bar and 
T~=~700~K, convolved to a resolving power of 1000, and is courtesy of 
R.S. Freedman, D. Schwenke, and D. Saumon.}

\end{document}